\documentclass[fleqn,12pt,twoside]{article}
\usepackage{espcrc1}

\usepackage{graphicx}
\usepackage[figuresright]{rotating}

\newcommand{\AmS}{{\protect\the\textfont2
  A\kern-.1667em\lower.5ex\hbox{M}\kern-.125emS}}

\title {Evaluation of the isospin  asymmetry of the nucleon 
structure functions with CLAS++}

\author{S. I. Alekhin \address{Institute of High Energy Physcics, 
142281  Protvino, Russia} \thanks{Supported by RFBI Grant N 03-02-17177},  
                A. L. Kataev \address{Institute for Nuclear Research of the 
Academy of Sciences of Russia, 117312 Moscow, Russia} \thanks{ Supported by 
RFBI Grants N 03-02-17047, 03-02-17177},  
S. A. Kulagin$^{\rm\ b\ *}$
and M. V. Osipenko \address{INFN, Sezione di Genova, I-16146, Genova, Italy 
and 
Skobeltsyn Institute of Nuclear Physics 119992 Moscow,  
Moscow State University, 119899 Moscow, Russia}}  
  
\begin{document}

\maketitle

\begin{abstract}
The possibility to estimate the isospin symmetry breaking effects
in the non-perturbative part of $F_2$ structure function of the charged  $ l N$  
deep-inelastic scattering  data, which will provide  CLAS++ detector of the
upgraded TJNAF machine  at $Q^2\approx 2$ GeV$^2$, is discussed.
The problems of the Gottfried sum rule extraction are also considered.
\end{abstract}

\section{INTRODUCTION}
First experimental evidence of the isospin asymmetry in 
structure functions (SFs) $F_2^{lp}(x,Q^2)$ and $F_2^{ln}(x,Q^2)$
came from earlier
studies of deep-inelastic scattering (DIS) at SLAC.
The Gottfried integral was extracted from the NMC measurement at CERN \cite{NMC}
\begin{equation}
I_{\rm GSR}(Q^2=4~{\rm GeV^2})=\int_0^1\frac{dx}{x}\bigg[F_2^{lp}-F_2^{ln}\bigg]
=0.235 \pm 0.026~~~~~ .
\label{GSR}
\end{equation}
This result is significantly different from the prediction of
the quark-parton model, $1/3$.  It is not possible to describe this
difference neither by the order $O(\alpha_s^2)$ perturbative QCD
corrections, nor by a twist-4 non-perturbative $1/Q^2$ contribution
\cite{BKM}.  These experimental results caused the discussion of the
isospin symmetry breaking effects (see, e.g., \cite{Kumano,review}). It
was realized, that in order to understand (\ref{GSR}) one needs to
assume a light-quark asymmetry of the nucleon see,
$\overline{u}(x)<\overline{d}(x)$, which has a non-perturbative
origin.  This concept found an additional experimental support in the
analysis of Drell-Yan process and the semi-inclusive DIS
\cite{review}.

However, it must be commented that the interpretation of the NMC
result (\ref{GSR}) can be affected by both the uncertainties because
of nuclear corrections (since the deuterium was used as effective
neutron target) and low-$x$ extrapolation. These problems were not
adequately addressed in Ref.\cite{NMC}.

\begin{figure}[!h]
\begin{center}
\includegraphics[bb=2cm 4cm 22cm 24cm, scale=0.35]{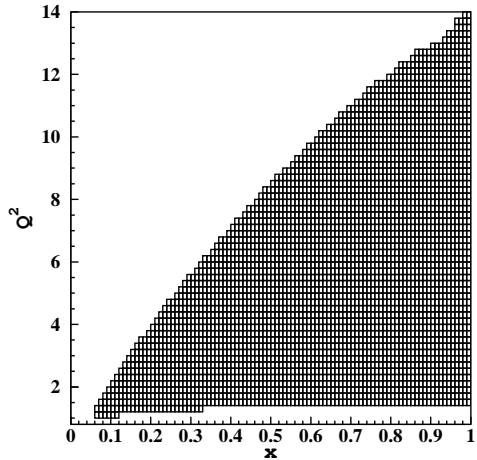}
\vspace{-1.5cm}\caption{\label{kinematics}
The shaded area shows the 
kinematics accessible in the inclusive electron scattering experiment with CLAS++
in the $(Q^2,x)$ plane.}
\end{center}
\end{figure}

A more sophisticated extraction of the isospin asymmetry from DIS
requires both the improvements in the accuracy of the measurements and the
extension of the kinematic region of data. In this respect we note
that the planned experiments after TJNAF upgrade to 12 ${\rm GeV}$
beam energy will significantly improve the accuracy of the measurement
of $F_2$ in the region of intermediate $Q^2$. The kinematic region
accessible with one of its detectors, CLAS++, is shown in Fig.~\ref{kinematics}.
It can be seen that $x=0.1$ is the lowest possible value of $x$ for DIS events
($Q^2>2$~GeV$^2$).
\vspace{-1.2cm}%
\vspace{-1.1cm}

Unfortunately the region $x<0.1$, which is crucial for the
evaluation of the Gottfried sum rule, can not be directly probed at
this machine. For this reason in what follows we discuss two
possibilities to study the isospin asymmetries using data from upgraded
TJNAF machine.

\section{ISOSPIN ASYMMETRY IN STRUCTURE FUNCTIONS} 

\begin{figure}[htb]
\begin{center}
\includegraphics[width=0.5\textwidth]{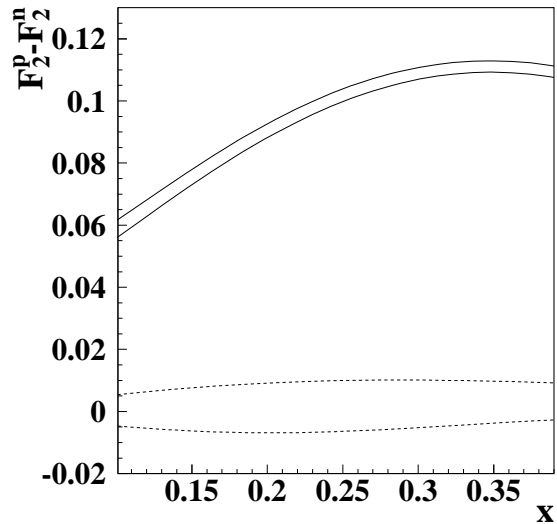}
\vspace{-0.9cm}%
\caption{\label{fig:f2p-n} 
The bands for the leading-twist (solid) and the high-twist (dashes)
contributions to the isospin asymmetry of the structure 
function $F_2$ at  $Q^2=2~\rm {GeV}^2$}. 
\end{center}
\end{figure}

One possible way to study the isospin asymmetries is to perform a
combined analysis of (future) TJNAF data and NMC data at low $x$, or
to extrapolate future TJNAF data to the region of low $x$ using
different sets of parton distributions, e.g. the one of
Ref.\cite{Alekhin:2002fv}.
Figure~\ref{fig:f2p-n} shows the difference
$F_2^p-F_2^n$, extracted from  the Next-to-Next-to-Leading order QCD 
analysis of the world experimental data on the charged lepton-nucleon DIS
cross-section \cite{AKL}.
Two solid lines give a band of the isospin asymmetry for the 
leading twist term and the dashed lines give the band accounting for the  
twist-4 term. The latter one turned out to be small, like in 
the first moment Gottfried sum rule \cite{BKM} and for higher moments 
as well  \cite{AKL}. However, the 
dominant uncertainty in the isospin 
asymmetry is given by non-perturbative effects. New   
CLAS++ data will give a better constraint on the 
twist-4 term in the isospin symmetry breaking effects in DIS
in the non-resonance region, which for $Q^2=2$ GeV$^2$ corresponds   
to  $x\leq 0.5$. This will allow
a more sophisticated treatment of higher-twist terms in the analysis of $F_2$ 
data, similar to those in Refs.~\cite{osipenko,AKL}.

The experimental methods of TJNAF measurement of the neutron structure 
$F_2$ from the deuteron target allows to eliminate 
a certain type of nuclear corrections by using the proton recoil detector. 
Such a recoil
detector for CLAS is now under construction at Jefferson 
Lab~\cite{Bonus}. New
hardware will not affect the accessible in CLAS kinematics, which
due to almost 4$\pi$ acceptance permits to measure the inclusive
cross section simultaneously in a wide region
of $x$ and $Q^2$ \cite{osipenko}. Therefore,
after TJNAF upgrade to 12 GeV beam energy, a combination of two measurements --
the measurement of the neutron $F_2$ from the deuteron target using
the proton recoil detector and the measurement of the proton $F_2$ from the 
hydrogen target --
will allow the extraction of isospin symmetry breaking effects.
The measurement of the neutron structure function is already planed
at CLAS, while the extraction of the proton structure function $F_2$
does not require an additional beam time and can be performed within
an analysis of electron run data collected during other experiments,
as it was done in~\cite{osipenko}.
\vspace{-1.5cm}%
\vspace{-2.1cm}

\section{THE GOTTFRIED SUM RULE} 

\begin{figure}[htb]
\begin{center}
\includegraphics[bb=2cm 4cm 22cm 24cm, scale=0.35]{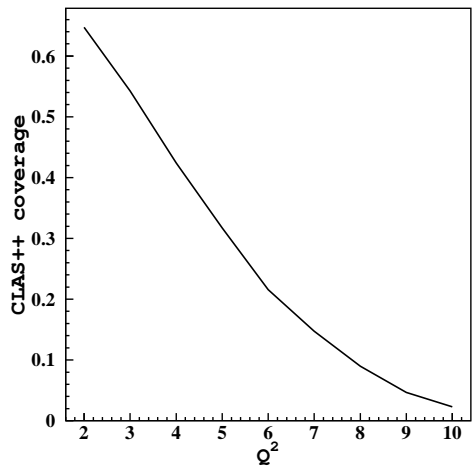}
\vspace{-1.2cm}\caption{\label{percentage} 
The share of the Gottfried sum rule accessible 
at CLAS++ data
calculated using the parameterizations of $F_2^p$ and $F_2^d$ \cite{Bodek,Milsztajn}
and the kinematics of Fig.~\ref{kinematics}.
} 
\end{center}
\end{figure}

The share of the Gottfried sum rule which is covered 
by CLAS++ data, according to kinematics shown at Fig.~\ref{kinematics}, is presented in
Fig.~\ref{percentage}. 
This estimate is  based on parameterizations of $F_2^{ep}$ 
and $F_2^{eD}$ from Refs.~\cite{Bodek,Milsztajn}. 
In the interval of $Q^2$ from 2 to 3.5 GeV$^2$
CLAS++ data can provide more than a half of the value of the Gottfried integral, assuming
the $x$ dependence of the proton and neutron structure functions is given by the 
parameterizations of
Ref.~\cite{Bodek,Milsztajn}. These parameterizations are quite reliable and
give the value of
Gottfried integral 0.2314 at $Q^2=4$ GeV$^2$ which is very close 
to the
value 0.235$\pm$0.026, obtained  at this $Q^2$ by 
NMC collaboration~\cite{NMC}.

\begin{figure}[!h]
\begin{center}
\includegraphics[width=0.5\textwidth]{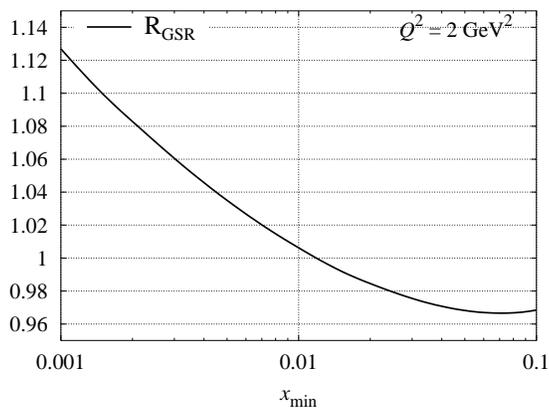}
\vspace{-1cm}\caption{\label{fig:nucgsr} 
Ratio 
(\ref{rgsr}) 
as a function of the cut $x_{\rm min}$ computed at $Q^2=2$\,GeV$^2$.
} 
\end{center}
\end{figure}

However, it is obvious, that
even at $Q^2=2$ GeV$^2$ the major uncertainties  of the 
measurement of the Gottfried sum rule at CLAS++ come from the extrapolation to
the region $x<0.1$. Moreover, even the uncertainties of the  well-known NMC 
result (\ref{GSR}) may be underestimated because of nuclear effects. 
In order to illustrate the integral strength of nuclear effects we have
calculated the ratio 
\begin{eqnarray} \label{rgsr} 
R_{\rm GSR} &=&
\frac{\int_{x_{\rm min}}^{x_{\rm max}}\,dx\left(2F_2^p(x,Q^2) - F_2^D(x,Q^2)\right)/x}
     {\int_{x_{\rm min}}^{x_{\rm max}}\,dx\left(F_2^p(x,Q^2) - F_2^n(x,Q^2)\right)/x}~. 
\end{eqnarray} 
In Eq.(\ref{rgsr}) $F_2^D$ is the deuteron structure function and the
integration is taken over the interval of Bjorken $x$ which is linked to
experimental conditions. In the absence of nuclear corrections 
$F_2^D=F_2^p +F_2^n$ that gives $R_{\rm GSR}=1$. The effect of 
nuclear corrections on the
ratio $R_{\rm GSR}$ as a function of the cut $x_{\rm min}$ is shown at 
Fig.~\ref{fig:nucgsr}.
In this calculation we used the model of the deuteron structure
function of Ref.\cite{Kulagin:2004ie}, the proton and the neutron structure
functions were calculated using the PDFs of Ref. \cite{Alekhin:2002fv}
and the upper limit of 
integration in (\ref{rgsr}) was fixed to $x_{\rm max}=0.4$. 
We observe about 3\% (negative) nuclear correction at $x_{\rm min}=0.1$ 
and at fixed $Q^2=2$\,GeV$^2$. However, 
the nuclear correction becomes positive and 
its magnitude rises as $x_{\rm min}$ decreases. We remark
that the rise of the magnitude of nuclear correction in this region is 
because $F_2^D < 2F_2^p$ at small $x$ 
(this inequality was verified by the E665 data \cite{E665dp}) and because of
the factor $1/x$ in the integrands in Eq.(\ref{rgsr}). 
\vspace{-1.5cm}%
\vspace{-1.1cm}
\vspace{-1.1cm}
\vspace{-0.5cm}

The Gottfried integral can be extracted from hydrogen and deuterium data as
\begin{equation} \label{gsr:p-d}
I_{\rm GSR}^{exp} = \int \frac{dx}{x} \left(
2F_2^p(x,Q^2) - \frac{F_2^D(x,Q^2)}{R_2(x,Q^2;D/N)}
\right)~~~,
\end{equation}
where the function $R_2(D/N)=F_2^D/(F_2^p+F_2^n)$ provides a correction 
for nuclear effects. 
An accurate model of this function has recently become available from the 
analysis of Ref.\cite{Kulagin:2004ie}.

Note, however,
even if the effects of nuclear corrections will be fixed 
the uncertainty of extrapolation to low $x$ using different  
modern sets of parton distributions of Refs.
\cite{Alekhin:2002fv}, \cite{Martin:2002aw},
\cite{Pumplin:2002vw} persists.

To conclude, the extraction of the Gottfried sum rule is difficult, but 
rather interesting problem.
Its studies will help to illuminate the effect
of light-quark flavour asymmetry $\overline{u}(x)-\overline{d}(x)$
as well as to test different extractions of 
$\overline{u}(x)$ and $\overline{d}(x)$ 
at low $x$ and clarify the reasons for these differences.

\section{ACKNOWLEDGEMENTS}
This contribution is the main updated part of the talk 
of one of us  
(ALK) at   Baryons 2004 Conference, Palaiseau. ALK is grateful to the members 
of Local Organizing Committee and in particular to M. Cuidal for hospitality 
in France. The research  of SIA was also supported in part by the Russian 
Ministry of Science grant NSh 1695.2003.2.

\end{document}